\documentclass[conference, english, 10pt]{IEEEtran}
\IEEEoverridecommandlockouts

\usepackage{babel}
\usepackage[babel]{microtype}
\usepackage[babel]{csquotes}

\usepackage[T1]{fontenc}
\usepackage{textcomp}

\usepackage[svgnames]{xcolor}
\usepackage{graphicx}
\usepackage{tikz}
\usepackage{pgfplots}
\usepackage{subfigure}

\usepackage{tabularx}
\usepackage{booktabs}

\usepackage{amsmath}
\usepackage{amsfonts}
\usepackage{bbm}
\usepackage{amssymb}
\usepackage{amsthm}
\usepackage{bm}
\usepackage{siunitx}
\usepackage{mathtools}
\usepackage{dsfont}

\usepackage{algorithmic}
\usepackage{xparse}
\usepackage{flushend}

\usepackage[backend=biber, style=ieee, url=false, isbn=false, dashed=false, doi=true]{biblatex}

\bibliography{ref.bib}

\usepackage{hyperref}
\hypersetup{
	pdfauthor={},
	pdftitle={},
	bookmarksnumbered=true,
	colorlinks=true,
	allcolors=.,
	urlcolor=MidnightBlue,
}

\usepackage[acronyms,nomain,xindy,nowarn]{glossaries}
\makeglossaries
\newacronym{ao}{AO}{alternating optimization}
\newacronym{ap}{AP}{access point}

\newacronym{bcd}{BCD}{block coordinate descent}
\newacronym{bs}{BS}{base station}

\newacronym{csi}{CSI}{channel state information}
\newacronym{cf-mmimo}{CF mMIMO}{cell free massive MIMO}
\newacronym{cpu}{CPU}{central processing unit}

\newacronym{drl}{DRL}{deep reinforcement learning}
\newacronym{ddpg}{DDPG}{deep deterministic policy gradient}

\newacronym{ee}{EE}{energy efficiency}
\newacronym{fcn}{FCN}{fully convolutional network}

\newacronym{gnn}{GNN}{graph neural network}
\newacronym{iid}{i.i.d.}{independent and identically distributed}

\newacronym{los}{LoS}{line-of-sight}
\newacronym{lr}{LR}{learning rate}

\newacronym{mpc}{MPC}{multi-path component}
\newacronym{mrt}{MRT}{maximum ratio transmission}
\newacronym{mse}{MSE}{mean squared error}
\newacronym{mmse}{MMSE}{minimum mean squared error}
\newacronym{mimo}{mMIMO}{massive multiple-input-multiple-output}
\newacronym{ml}{ML}{machine learning}

\newacronym{nn}{NN}{neural network}
\newacronym{ris}{RIS}{reconfigurable intelligent surface}

\newacronym{snr}{SNR}{signal to noise ratio}
\newacronym{sinr}{SINR}{signal to interference noise ratio}
\newacronym{sdma}{SDMA}{space-division multiple access}
\newacronym{sca}{SCA}{sequential convex approximation}

\newacronym{tsnr}{TSNR}{transmit signal to noise ratio}
\newacronym{ue}{UE}{user equipment}
\newacronym{wmmse}{WMMSE}{weighted minimum mean squared error}
\newacronym{wsr}{WSR}{weighted sum rate}

\newacronym{zf}{ZF}{zero forcing}

\setacronymstyle{long-short}
\glsdisablehyper

\usetikzlibrary{positioning}
\usetikzlibrary{shapes.geometric, arrows, decorations.pathreplacing}
\usetikzlibrary{shapes.arrows}
\usetikzlibrary{3d}
\pgfplotsset{compat=newest}

\tikzstyle{block} = [rectangle, rounded corners, text width=2cm, text centered, draw=black]

\tikzstyle{layer} = [rectangle, rounded corners, minimum width=2.5cm, minimum height=.6cm, align=center, text centered, draw=black]
\makeatletter 
\tikzoption{canvas is xy plane at z}[]{%
  \def\tikz@plane@origin{\pgfpointxyz{0}{0}{#1}}%
  \def\tikz@plane@x{\pgfpointxyz{1}{0}{#1}}%
  \def\tikz@plane@y{\pgfpointxyz{0}{1}{#1}}%
  \tikz@canvas@is@plane
}
\makeatother
\NewDocumentCommand{\DrawCubes}{O {} m m m m m m}{%
    \def\XGridMin{#2}
    \def\XGridMax{#3}
    \def\YGridMin{#4}
    \def\YGridMax{#5}
    \def\ZGridMin{#6}
    \def\ZGridMax{#7}
    \begin{scope}[canvas is xy plane at z=\ZGridMax]
      \draw [#1] (\XGridMin,\YGridMin) grid (\XGridMax,\YGridMax);
    \end{scope}
    \begin{scope}[canvas is yz plane at x=\XGridMax]
      \draw [#1] (\YGridMin,\ZGridMin) grid (\YGridMax,\ZGridMax);
    \end{scope}
    \begin{scope}[canvas is xz plane at y=\YGridMax]
      \draw [#1] (\XGridMin,\ZGridMin) grid (\XGridMax,\ZGridMax);
    \end{scope}
}%

\newcommand{\wrt}{w.r.t}
\newcommand{\nn}{SINRnet}
\newcommand{\todo}[2][]{\ignorespaces
	\if\relax\detokenize{#1}\relax
	{\color{red}[TODO: #2]}%
	\else
	{\color{red}[TODO (#1): #2]}%
	\fi
}

\addto\extrasenglish{%

}
\usetikzlibrary{spy}

\begin{document}

\title{Energy-Efficient Power Allocation in Cell-Free Massive MIMO via Graph Neural Networks 
\thanks{The work is partly supported by the Federal Ministry of Education and Research Germany (BMBF) as part of the 6G Research and Innovation Cluster 6G-RIC under Grant 16KISK031.}
}

\author{%
\IEEEauthorblockN{%
Ramprasad Raghunath, Bile Peng, Eduard A. Jorswieck}
\IEEEauthorblockA{Institute for Communications Technology, Technische Universit\"at Braunschweig, Germany}

Email: \{r.raghunath, b.peng, e.jorswieck\}@tu-braunschweig.de
}

\maketitle

\begin{abstract}
\Gls{cf-mmimo} systems are a promising solution to enhance the performance in 6G wireless networks. 
Its distributed nature of the architecture makes it highly reliable, provides sufficient coverage and allows higher performance than cellular networks. 
\Gls{ee} is an important metric that reduces the operating costs and also better for the environment.
In this work, we optimize the downlink \gls{ee} performance with \gls{mrt} precoding and power allocation.
Our aim is to achieve a less complex, distributed and scalable solution. 
To achieve this, we apply unsupervised \gls{ml} with permutation equivariant architecture and use a non-convex objective function with multiple local optima.
We compare the performance with the centralized and computationally expensive \gls{sca}.
The results indicate that the proposed approach can outperform the baseline with significantly less computation time.
\end{abstract}


\glsresetall

\section{Introduction}
\label{sec:rep}
\Gls{cf-mmimo} uses many spatially distributed\glspl{ap} to serve \glspl{ue} and overcome the performance bottleneck in the cellular networks.
From the \gls{ue}'s point of view the cell boundaries disappear and thus obtain seamless and uniform coverage even with high mobility~\cite{Ammar2022}.
The distributed nature of \gls{cf-mmimo} can have many advantages.
For example,
the \gls{ue} is more likely to have a line-of-sight channel and
\gls{cf-mmimo} enables distributed processing useful in serving large number of \glspl{ue}.
There have been many works to improve the performance of \gls{cf-mmimo} system.
\cite{jin2020spectral} proposes to maximize the spectral efficiency and \gls{ee} with successive approximation.
\cite{nguyen2017energy} reduces the complexity of the problem with a \gls{zf} precoder design.
In~\cite{Zhao2020} a deep learning method is proposed to approximate a high complexity algorithm for max-min power control.
\cite{DAndrea2019} performs uplink power control for sum-rate and max-min rate optimization using deep learning. 
\cite{Zhang2021} uses unsupervised deep learning to optimize max-min, max-product and max-sum-rate optimization with the same \gls{nn} structure.
\cite{Luo2022} proposes a \gls{drl} approach with \gls{ddpg} framework to address the downlink max-min power control problem and then extend the framework to max-sum and max-product power control problems of  \gls{cf-mmimo}.
There are several other works in the literature that consider max-min power control technique~\cite{Ngo2017,Nayebi2017,Chakraborty2019}.

The analytical methods found in the literature are either computationally expensive and/or must be implemented in a centralized way.
The supervised \gls{ml} approaches have an expensive training routine and their architecture is not scalable or symmetric.
In this work, we use the \gls{gnn} with unsupervised learning framework to optimize the downlink power control to maximize the \gls{ee} of a \gls{cf-mmimo} system with \gls{mrt} precoding. We choose the unsupervised learning framework because there is no need to prepare labels for the training process which involves solving the non-convex problem which is computationally expensive.

A \gls{gnn} is an optimizable transformation on all attributes of the graph (nodes, edges, global-context) that preserves graph symmetries (permutation invariances)~\cite{SanchezLengeling2021}. 
\gls{gnn} architecture recieved a lot of interest in the past few years.
We think that this architecture fits best to \gls{cf-mmimo} in terms of distributed computing.
\cite{Salaun2022} proposes a \gls{gnn} to solve the downlink max-min power control problem in a supervised learning framework.
It is also shown in~\cite{Shen2023} that \gls{gnn} are more sample efficient and better at solving problems in wireless communication as compared to standard \gls{nn}.
The proposed framework has several advantages over analytical optimization.
First, according to the universal approximation theorem~\cite{Hornik1989}, deep \glspl{nn} can approximate any continuous function when trained properly, which ensures better performance than methods with weaker approximations.
Moreover, the data flow and processing happens in a distributed way unlike other analytical methods which need all the data in a \gls{cpu} which requires significant signalling and front-haul capacity.
It is also of interest that similar design principles are followed in \gls{cf-mmimo}, i.e., distributed architecture.
This allows us to realize a fully distributed optimization problem without the need of a \gls{cpu}.
In addition, the \gls{gnn} architecture is symmetric and suitable for power control problem (explained in \autoref{sec:gnn}).
Our contributions in this work are as follows:
\begin{itemize}
    \item We develop a custom \gls{gnn} with simple message passing and nested \nn~\cite{Peng2023}. 
    \item We use an unsupervised and non-convex loss function that helps in achieving a near optimal performance in an efficient way.
    \item We train the network and compare its performance to the state of the art \gls{sca} to evaluate its performance.
\end{itemize}
The numerical assessment show that we can reduce the complexity of power allocation significantly and also achieve better \gls{ee} performance with the proposed approach.

Notation: Boldface uppercase and lowercase letters denote matrices and column vectors, respectively. 
$(\cdot)^T$, $(\cdot)^*$, and $(\cdot)^H$ denote transpose, conjugate, and conjugate transpose operations, respectively. 
$\mathbbm{1}$ denotes an vector of ones, and $|\cdot|$ denotes the norm operation. 

\section{System Model and Problem Formulation}
\subsection{Cell-free Massive MIMO}
\begin{figure}
    \centering
    \resizebox{.5\linewidth}{!}{
    \input{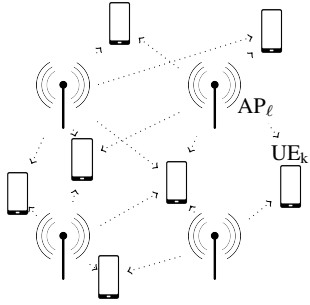}}
    \caption{Cell-free massive MIMO setup}
    \label{fig:cell-free}
\end{figure}

We consider a cell-free network~\cite{Bjornson2020} with $K$ single antenna \glspl{ue} and $L$ \glspl{ap} with $N$ antennas each as shown in~\autoref{fig:cell-free}.
The \glspl{ap} are in a grid and the \glspl{ue} are uniformly distributed over a square coverage area.
The \glspl{ap} are allowed to communicate with each other over front-haul to enable non-coherent joint transmission to the \glspl{ue}~\cite{Vu2020}.

We assume that \glspl{ap} have full \gls{csi} of the channels with channel gains above a certain threshold and each \gls{ue} can be served by only \glspl{ap} which fulfill this threshold, this means that \gls{csi} is not available globally.
In addition we also assume that \glspl{ap} use \gls{mrt} precoding for the downlink transmission.
\Gls{ap} $l$ assigns transmit power $p_{kl}$ to \gls{ue} $k$.

\subsection{Problem Formulation}
Our objective is to maximize the sum \gls{ee}, which is defined as the sum of ratios between data rate and power consumption of all links.
The problem is formulated as
\begin{equation}
\begin{aligned}
    \max_{\mathbf{P}} \quad & J(\mathbf{P}) = \sum_{k=1}^K \frac{\log\left(1 + \frac{\sum_{l=1}^L h_{l,kk} p_{kl}}{\sigma^2 + \sum_{j \neq k} \sum_{l=1}^L h_{l,kj} p_{jl}}\right)}{\mu \sum_{l=1}^L p_{kl} + P_c}\\
    \text{s.t.} \quad & p_{kl} \geq 0 \quad \text{ for } k = 1, 2, \dots, K, \text{ and } l = 1,2, \dots, L,
\end{aligned}
\label{eq:problem}
\end{equation}
where $\mathbf{P} \in \mathbb{R}^{K \times L}$ is the transmit power matrix between all \glspl{ap} and \glspl{ue}. 
$\mathbf{h}_{l,k}$ is the effective channel vector between \gls{ue} $k$ and \gls{ap} $l$ with \gls{mrt} precoding, 
\begin{equation}
    \label{eq:mrt}
    h_{l,kj} = \Big| \mathbf{h}_{l,k}^H \cdot \frac{\mathbf{h}_{l,j}}{\|\mathbf{h}_{l,j}\|} \Big|^2,
\end{equation}
for \gls{ue} $k$ and the transmitter $j$.
Therefore $h_{l,kk}$ is the received gain of user $k$ and $h_{l,kj}$, $j \neq k$ is the interference from \gls{ue} $j$. 
Since this is a distributed architecture and each \gls{ap} has its features, we denote its elements as $h_{kj}$ for \gls{ap} (node) $l$ in the rest of the paper.
$\sigma^2$ is the noise power at the receiver.
$\mu$ is the inefficiency of the power amplifier and $P_c$ is the static power consumption of the system. 

\begin{figure}[htbp]
   \centering
   \begin{tikzpicture}
\draw[step=0.5cm] (0, 0) grid (1.5, 1.5);
\node (ap1) [anchor=east] at (0, 0.25) {UE 3};
\node (ap2) [anchor=east] at (0, 0.75) {UE 2};
\node (ap3) [anchor=east] at (0, 1.25) {UE 1};
\node (ue1) [rotate=-90, anchor=east] at (0.25, 1.5) {AP 1};
\node (ue2) [rotate=-90, anchor=east] at (0.75, 1.5) {AP 2};
\node (ue3) [rotate=-90, anchor=east] at (1.25, 1.5) {AP 3};

\draw[->] (0.75, -0.05) -- (0.75, -0.45) {};

\draw[step=0.5cm] (0, -1) grid (1.5, -0.5);
\node (p1) [anchor=south] at (0.25, -1) {$p_1$};
\node (p2) [anchor=south] at (0.75, -1) {$p_2$};
\node (p3) [anchor=south] at (1.25, -1) {$p_3$};

\newcommand\shift{4}

\draw[step=0.5cm] (0 + \shift - 0.00001, 0) grid (1.5+ \shift, 1.5);
\node (ap1) [anchor=east] at (0 + \shift, 0.25) {\color{DodgerBlue} UE 2};
\node (ap2) [anchor=east] at (0 + \shift, 0.75) {\color{DodgerBlue} UE 3};
\node (ap3) [anchor=east] at (0 + \shift, 1.25) {UE 1};
\node (ue1) [rotate=-90, anchor=east] at (0.25 + \shift, 1.5) {AP 1};
\node (ue2) [rotate=-90, anchor=east] at (0.75 + \shift, 1.5) {\color{DodgerBlue} AP 3};
\node (ue3) [rotate=-90, anchor=east] at (1.25 + \shift, 1.5) {\color{DodgerBlue} AP 2};

\draw[->] (0.75 + \shift, -0.05) -- (0.75 + \shift, -0.45) {};

\draw[step=0.5cm] (0 + \shift - 0.00001, -1) grid (1.5 + \shift, -0.5);
\node (p1) [anchor=south] at (0.25 + \shift, -1) {$p_1$};
\node (p2) [anchor=south] at (0.75 + \shift, -1) {\color{DodgerBlue} $p_3$};
\node (p3) [anchor=south] at (1.25 + \shift, -1) {\color{DodgerBlue} $p_2$};

\end{tikzpicture}%
   \caption{Illustration of permutation-equivariance.}
   \label{fig:equivariance}
\end{figure}
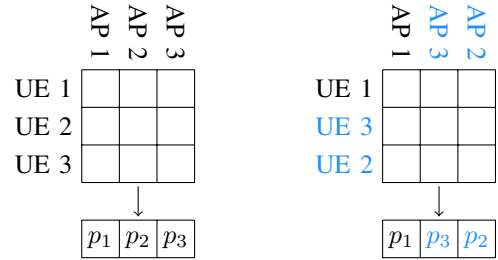

We note that the order of \glspl{ue} served by \gls{ap} $l$ in \eqref{eq:problem} should be arbitrary.
If we permute the order in the input,
the output order should be permuted in the same way
as~\autoref{fig:equivariance} shows.
This property is named permutation-equivariance.

The \gls{ue} permutation equivariance is taken care by the \nn architecture and the \gls{ap} permutation equivariance is ensured by the symmetric architecture of \gls{gnn}. 
\section{Non-convex Objective Function and Support Regularization}
\subsection{Unsupervised Learning Framework}
\label{subsec:ulf}
Given a \gls{csi} $\mathbf{H}$,
we look for a power allocation $\mathbf{P}$ that maximizes \gls{ee},
which is fully determined by $\mathbf{H}$ and $\mathbf{P}$ and can be written as
$J(\mathbf{H}, \mathbf{P})$.
We define a \gls{nn} $N_\lambda$, which is parameterized by $\lambda$ and
maps from $\mathbf{H}$
to $\mathbf{P}$,
i.e., $\mathbf{P} = N_\lambda(\mathbf{H}).$
We can write the objective as $ J(\mathbf{H}, \mathbf{P})=J(\mathbf{H}, N_\lambda(\mathbf{H})),$ 
note that the equation emphasizes that $J$ depends on $\lambda$ given $\mathbf{H}$.
We collect massive data of $\mathbf{H}$ in a training set $\mathcal{D}$
and formulate the unsupervised \gls{ml} problem as
\begin{equation}
    \max_\lambda \sum_{\mathbf{H} \in \mathcal{D}} J(\mathbf{H}, N_\lambda(\mathbf{H})).
    \label{eq:ml}
\end{equation}
This way, we optimize $N_\lambda$ for any $\mathbf{H} \in \mathcal{D}$ (\emph{training}).
If the data set is general enough,
a data sample $\mathbf{H}' \notin \mathcal{D}$
can also be mapped to a good action (\emph{testing}),
like a human can use experience to solve new problems of the same type\footnote{A complete retraining is only required when the input states are fundamentally changed, e.g., change of carrier frequency in wireless context.}~\cite{yu2022role}.
This approach has been successfully applied in~\cite{Peng2023} on interference network.
This framework contains the complexity in training and helps keep the application fairly simple. 
When compared with analytical methods like \gls{sca} which do not particularly have a distinction between learning/application.

\subsection{Non-convex Objective Function}
\label{subsec:non-convex}
The objective in \eqref{eq:problem} is a non-convex function, which means there might exist multiple local optima.
If \gls{nn} uses gradient methods to optimize its parameters \wrt~the objective (loss) function, 
the initial value of the solution decides the convergence.
If $\mathbf{p}_0$ is initialized near a poor local optimum, 
the solution might converge to this poor local optimum.

Since the converged local optimum depends on the initialization of $\mathbf{p}$
and we do not know the correct initialization before the optimization,
we do not assume $\mathbf{p}$ is deterministic,
but let $p$ be uniformly distributed in the feasible region i.e., $\mathbf{p} \sim \mathcal{U}(\mathbf{a}, \mathbf{b})$ where $\mathbf{a}$ and $\mathbf{b}$ are the upper and the lower bound of the support of the distribution and are output of the \gls{gnn} given $\mathbf{H}$.
Throughout the process of optimization, the support of the distribution is narrowed until convergence to the global optimum is achieved. 

Since we do not use a deterministic value for the optimization variable $\mathbf{p}$, but a random variable sampled from a distribution, the objective function has to be modified to account for the variance.
Therefore, the expectation of $J$ defined as
\begin{equation}
    K = \mathbb{E}_{\mathbf{p} \sim \mathcal{U}(\mathbf{a}, \mathbf{b})}(J\left(\mathbf{p})\right)
    \label{eq:objective1}
\end{equation}
is considered as the objective.
We can use the mean of multiple \gls{iid} random variables to approximate the expectation in \eqref{eq:objective1}.
However, the gradient of a random variable cannot be computed and gradient ascent cannot be applied to maximize $K$ as a result.
Hence, the reparametrization trick~\cite{kingma2013} is applied to make sure that $\mathbf{p} \sim \mathcal{U}(0,1)$ is sampled from a fixed distribution and~\eqref{eq:objective1} can be rewritten as 
\begin{equation}
    K = \mathbb{E}_{\mathbf{p} \sim \mathcal{U}(\mathbf{0}, \mathbf{1})}(J\left(\mathbf{a} + (\mathbf{b} - \mathbf{a}) \odot \mathbf{p})\right),
    \label{eq:objective1.1}
\end{equation}
where $\odot$ denotes the element-wise product. And $\mathbf{a} + (\mathbf{b} - \mathbf{a}) \odot \mathbf{p} \sim \mathcal{U}(\mathbf{a}, \mathbf{b})$ if $\mathbf{p} \sim \mathcal{U}(\mathbf{0}, \mathbf{1})$.
Due to the reparametrization of the random variable, it is now possible to replace $J$ with $K$ in \eqref{eq:ml}, compute $\nabla_{\mathbf{a}}K$ and $\nabla_{\mathbf{b}}K$ and perform a gradient ascent step to improve $K$.

An illustration of the optimization in one dimension is shown in Figure 3 in~\cite{Peng2023}.
In this case, $a$ and $b$ cannot converge if we use \eqref{eq:objective1.1} as the objective function.
Therefore, we introduce a penalty term for support regularization in the objective as an incentive to reduce the support of the distribution.
Since we consider a multivariate uniform distribution, the penalty function considered is
\begin{equation}
    \psi(\mathbf{a},\mathbf{b}) = \mathbbm{1}^T \cdot (\mathbf{b} - \mathbf{a} ).
    \label{eq:support}
\end{equation}

The combined objective using \eqref{eq:objective1.1} and the penalty in \eqref{eq:support} is given as
\begin{equation}
\begin{aligned}
    L = & \mathbb{E}_{\mathbf{p} \sim \mathcal{U}(\mathbf{0}, \mathbf{1})} J(\mathbf{a} + (\mathbf{b}
    - \mathbf{a})\odot\mathbf{p})
     - \kappa \psi(\mathbf{a}, \mathbf{b}),
\end{aligned}
\label{eq:objective2}
\end{equation}
where $\kappa$ is the support regularization coefficient, a tunable parameter during the optimization process.
$\nabla_{\mathbf{a}} L$ and $\nabla_{\mathbf{b}} L$ are computed and gradient ascent is performed to update $\mathbf{a}$ and $\mathbf{b}$ until the support of $\mathcal{U}(\mathbf{a},\mathbf{b})$ is small enough to eradicate the stochasticity of the solution (making it practically deterministic).

From \eqref{eq:objective2} it is clear that $\kappa$ is a weighting factor on how much the penalty contributes to the global objective. 
So, if $\kappa$ is increased, the optimizer prioritizes minimizing the support, discarding local optima in the process.
Also since local optima have lower value than the global optimum, small increments to $\kappa$ result in pruning local optima gradually.
After getting rid of local optima, $\kappa$ decreases to keep the global optimum in the support of $[\mathbf{a},\mathbf{b}]$.
Compared to the conventional gradient optimization method,
in which the converged local optimum depends on the initialization,
the proposed method initializes the optimization variable as a random variable over the entire feasible region.
Therefore, it covers all possibilities at the beginning.
By improving the expectation of the objective \eqref{eq:objective1}
aided by the support regularization \eqref{eq:support},
the proposed method has a significantly higher possibility to approach the global optimum.

\subsection{Support Regularization}
Support regularization plays a crucial role in the success of the training because it helps to escape the local optima while ensuring global optimum is still in the support of $[\mathbf{a},\mathbf{b}]$.
This idea was introduced as entropy regularization in~\cite{Peng2023} for \gls{ee} maximization under power constraints for interference networks.
It is a common practice to change the problem into dual form to optimize the regularization factor~\cite{haarnoja2018soft}.
However, it is very difficult to obtain the infimum of the Lagrangian of our complicated objective function (unlike the canonical definition of the dual problem),
which would be a fatal disadvantage in the considered problem.
A simple heuristic approach was proposed in~\cite{Peng2023} to tune $\kappa$. 
Consider $\kappa_i$ in iteration~$i$ of the gradient ascent optimization.
$\kappa_i = 0$ for $i \leq h$,
where $h$ is a hyper-parameter for constant $\kappa$.
For $i > h$,
$\kappa$ is computed as
\begin{equation}
    \kappa_{i + 1} =
    \begin{cases}
    \kappa_i + \Delta\kappa & \text{if}\; \sum_{j = 1}^{h} \psi_{i - j} / h \leq \psi_i \\
    \max(0, \kappa_t - \Delta\kappa / 2) & \text{otherwise,}\\
    \end{cases}
    \label{eq:updating_kappa}
\end{equation}
where $\Delta\kappa$ is a constant small learning rate.
The intuition behind \eqref{eq:updating_kappa} is that we carefully increase $\kappa$ if the support does not reduce (case~1)
and decrease $\kappa$ otherwise such that $\kappa$ is not too big to make $\mathbf{a}$ or $\mathbf{b}$ cross the global optimum (case~2).
\section{GNN with nested \nn}
\label{sec:gnn}
This section presents the architecture of the \gls{gnn} developed in this work. 
The standard \gls{gnn} with message passing scheme~\cite{Fey2019} is represented as 
\begin{equation}
\label{eq:message-passing-og}
    \mathbf{x}_i^{(n)} = \gamma^{(n)} \left( \mathbf{x}_i^{(n-1)}, \bigoplus_{j \in \mathcal{N}(i)} \, \phi^{(n)}\left(\mathbf{x}_i^{(n-1)}, \mathbf{x}_j^{(n-1)},\mathbf{e}_{j,i}\right) \right),
\end{equation}
where $n$ denotes the current layer, $i$ denotes the current node, $j$ denotes the neighboring node, $e_{j,i}$ denotes the edge feature from node $j$ to $i$ and $x_i$, $x_j$ denotes the node feature of $i$ and $j$.
$\phi$ is \gls{nn} for message passing.
$\bigoplus$ denotes the aggregation function.
$\gamma$ is the feature update function at each node.

In the context of \gls{cf-mmimo}, each \gls{ap} can be represented as a node, the front haul connection between these \glspl{ap} can be represented as edge connecting the nodes and \gls{csi} corresponding to the \gls{ue} they serve is its node and edge feature.
It is important to note that in the standard form, $\phi$ accepts $x_i$ as input, this means node $i$ has to communicate with all its neighboring nodes to pass the message, node $j$ then processes this information and passes this message back to node $i$ which is then aggregated and passed onto $\gamma$.
This results in significant front-haul communication.
Therefore, we model the network with a  neighborhood aggregation or message passing scheme as
\begin{equation}
    \label{eq:message-passing}
    \mathbf{x}_i^{(n)} = \gamma^{(n)} \left( \mathbf{x}_i^{(n-1)}, \bigoplus_{j \in \mathcal{N}(i)} \, \phi^{(n)}\left( \mathbf{x}_j^{(n-1)},\mathbf{e}_{j,i}\right) \right).
\end{equation}

Since we use a custom architecture of \gls{gnn} (explained in \autoref{sec:gnn}), the message from each node is only computed once and broadcast to all the other nodes hence providing a significant advantage of reducing the computation and fronthaul communication.
In~\eqref{eq:message-passing}, the function $\phi$ is the edge processing network, $\gamma$ is the node processing network and we considered \nn~\cite{Peng2023} for both the functions.

The \nn{} is comprised of $V$ layers.
The channels between \gls{ue}~{k} and \gls{ap} (node)~$l$ are represented by $h_{kj}$ as the input of layer~$1$.
The feature matrix of $h_{kj}$ is given as $\mathbf{F}_{k'j',n}$ and it is the input to layer~$n$.  
For $n < V$ (i.e., for layers before the last layer), 
the output feature matrix of channel~$h_{lk}$ in layer~$n$ for node $l$ (i.e., the input feature in layer~$n + 1$) is computed as
\begin{equation}
\mathbf{F}^c_{h_{kj, n + 1}} = \sum_{h_{k'j'} \in \mathcal{S}^c(h_{kj})}\text{ReLU}(\mathbf{W}_n^c \mathbf{F}_{k'j', n} + \mathbf{b}_n^c) / |\mathcal{S}^c(h_{kj})|,
\label{eq:nic_layer}
\end{equation}
where $\mathbf{W}^c_n$ and $\mathbf{b}^c_n$ are the trainable weights and bias of layer~$n$ for category~$c$, respectively,
$|\mathcal{S}|$ denotes the cardinality of channels in category $\mathcal{S}$.
The channel categories are based on their position in the \gls{sinr} expression as follows:
\begin{enumerate}
    \item Category \num{1} is channel $h_{kk}$ which is the channel gains of the useful signals for \gls{ue} $k$. 
    \item Category \num{2} is channel $h_{kj}$ which is the interference channel between transmitter $j$ and user $k$.
    \item Category \num{3} is the interference channel between transmitter $k$ and \gls{ue} $j$ served by the \gls{ap}.
    \item Category \num{4} is the remaining channels between transmitter $j$ and \glspl{ue} $j$.
\end{enumerate}

\nn ~is suitable for $\phi$ and $\gamma$ because of its ability to include the domain knowledge and because this is a \gls{sinr} related power control, categorize the channels and also its permutation equivariant property~\cite{Peng2023}.

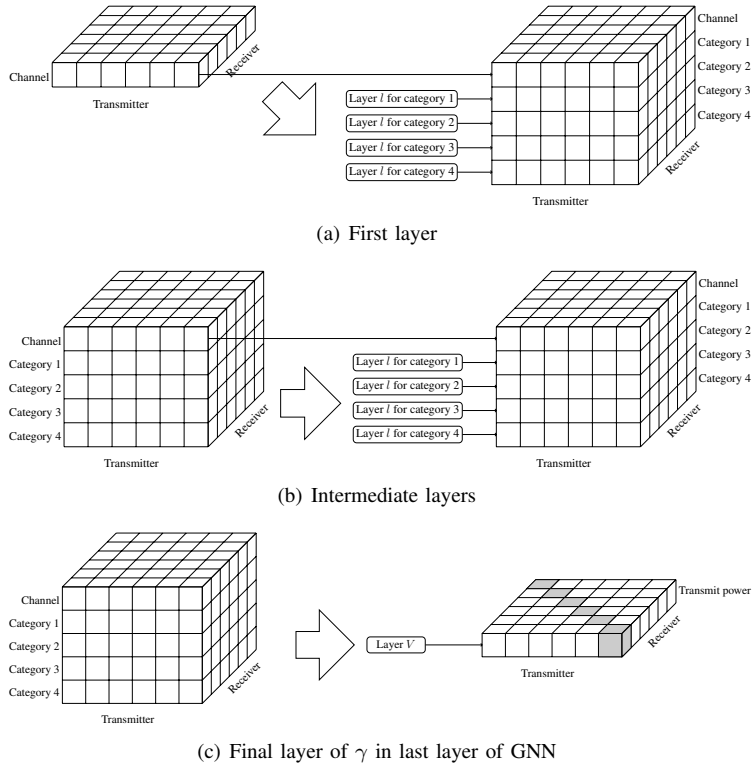
\begin{figure*}[htbp]
\centering
    \subfigure[First layer]{
    \resizebox{.55\linewidth}{!}{\begin{tikzpicture}
y={(0.5cm,0.25cm)},x={(0.5cm,-0.25cm)},z={(0cm,{veclen(0.5,0.25)*1cm})}
    ]
    \DrawCubes [step=10mm,thin]{0}{6}{4}{5}{0}{6}
    \node (width) [rectangle, yshift=1cm, xshift=.5cm,font=\Large] {Transmitter};
    \node (height) [rectangle, yshift=2.1cm, xshift=-2.3cm, anchor=east,font=\Large] {Channel};
    \node (channels) [rectangle, yshift=2.8cm, xshift=5.5cm,rotate=45,font=\Large] {Receiver};

\node[draw, single arrow,
              minimum height=25mm, minimum width=30mm,
              single arrow head extend=2mm,
              anchor=west, rotate=-45] at (6.7, 1.5) {};

\node (layer1) [layer, xshift=12cm, yshift=1.2cm,font=\Large] {Layer $l$ for category 1};
\node (layer2) [layer, below of=layer1, yshift=0cm,font=\Large] {Layer $l$ for category 2};
\node (layer3) [layer, below of=layer2, yshift=0cm,font=\Large] {Layer $l$ for category 3};
\node (layer4) [layer, below of=layer3, yshift=0cm,font=\Large] {Layer $l$ for category 4};

\draw [->] (3.7, 2.2) -- (15.7, 2.2);
\draw [->] (layer1.east) -- (15.7, 1.2);
\draw [->] (layer2.east) -- (15.7, 0.2);
\draw [->] (layer3.east) -- (15.7, -0.8);
\draw [->] (layer4.east) -- (15.7, -1.8);

    \DrawCubes [step=10mm,thin]{17.99}{24}{0}{5}{0}{6}
    \node (width) [rectangle, yshift=-3cm, xshift=18.5cm,font=\Large] {Transmitter};
    \node (height) [rectangle, yshift=4.5cm, xshift=24cm, anchor=west,font=\Large] {Channel};
    \node (height) [rectangle, yshift=3.5cm, xshift=24cm, anchor=west,font=\Large] {Category 1};
    \node (height) [rectangle, yshift=2.5cm, xshift=24cm, anchor=west,font=\Large] {Category 2};
    \node (height) [rectangle, yshift=1.5cm, xshift=24cm, anchor=west,font=\Large] {Category 3};
    \node (height) [rectangle, yshift=0.5cm, xshift=24cm, anchor=west,font=\Large] {Category 4};
    \node (channels) [rectangle, xshift=14cm, yshift=-1.2cm, xshift=9.5cm,rotate=45,font=\Large] {Receiver};
\end{tikzpicture}}}
    \subfigure[Intermediate layers]{
    \resizebox{.55\linewidth}{!}{    \begin{tikzpicture}
y={(0.5cm,0.25cm)},x={(0.5cm,-0.25cm)},z={(0cm,{veclen(0.5,0.25)*1cm})}
    ]
    \DrawCubes [step=10mm,thin]{0}{6}{0}{5}{0}{6}
    \node (width) [rectangle, yshift=-3cm, xshift=.5cm,font=\Large] {Transmitter};
    \node (height) [rectangle, yshift=2.1cm, xshift=-2.3cm, anchor=east,font=\Large] {Channel};
    \node (height) [rectangle, yshift=1.1cm, xshift=-2.3cm, anchor=east,font=\Large] {Category 1};
    \node (height) [rectangle, yshift=0.1cm, xshift=-2.3cm, anchor=east,font=\Large] {Category 2};
    \node (height) [rectangle, yshift=-0.9cm, xshift=-2.3cm, anchor=east,font=\Large] {Category 3};
    \node (height) [rectangle, yshift=-1.9cm, xshift=-2.3cm, anchor=east,font=\Large] {Category 4};
    \node (channels) [rectangle, yshift=-1.2cm, xshift=5.5cm,rotate=45,font=\Large] {Receiver};

\node[draw, single arrow,
              minimum height=25mm, minimum width=30mm,
              single arrow head extend=2mm,
              anchor=west] at (6.7, -0.5) {};

\node (layer1) [layer, xshift=12cm, yshift=1.2cm,font=\Large] {Layer $l$ for category 1};
\node (layer2) [layer, below of=layer1, yshift=0cm,font=\Large] {Layer $l$ for category 2};
\node (layer3) [layer, below of=layer2, yshift=0cm,font=\Large] {Layer $l$ for category 3};
\node (layer4) [layer, below of=layer3, yshift=0cm,font=\Large] {Layer $l$ for category 4};

\draw [->] (3.7, 2.2) -- (15.7, 2.2);
\draw [->] (layer1.east) -- (15.7, 1.2);
\draw [->] (layer2.east) -- (15.7, 0.2);
\draw [->] (layer3.east) -- (15.7, -0.8);
\draw [->] (layer4.east) -- (15.7, -1.8);

    \DrawCubes [step=10mm,thin]{17.99}{24}{0}{5}{0}{6}
    \node (width) [rectangle, yshift=-3cm, xshift=18.5cm,font=\Large] {Transmitter};
    \node (height) [rectangle, yshift=4.5cm, xshift=24cm, anchor=west,font=\Large] {Channel};
    \node (height) [rectangle, yshift=3.5cm, xshift=24cm, anchor=west,font=\Large] {Category 1};
    \node (height) [rectangle, yshift=2.5cm, xshift=24cm, anchor=west,font=\Large] {Category 2};
    \node (height) [rectangle, yshift=1.5cm, xshift=24cm, anchor=west,font=\Large] {Category 3};
    \node (height) [rectangle, yshift=0.5cm, xshift=24cm, anchor=west,font=\Large] {Category 4};
    \node (channels) [rectangle, xshift=14cm, yshift=-1.2cm, xshift=9.5cm,rotate=45,font=\Large] {Receiver};
\end{tikzpicture}}}
    \subfigure[Final layer of $\gamma$ in last layer of \gls{gnn}]{
    \resizebox{.55\linewidth}{!}{    \begin{tikzpicture}
y={(0.5cm,0.25cm)},x={(0.5cm,-0.25cm)},z={(0cm,{veclen(0.5,0.25)*1cm})}
    ]
    \DrawCubes [step=10mm,thin]{0}{6}{0}{5}{0}{6}
    \node (width) [rectangle, yshift=-3cm, xshift=.5cm,font=\Large] {Transmitter};
    \node (height) [rectangle, yshift=2.1cm, xshift=-2.3cm, anchor=east,font=\Large] {Channel};
    \node (height) [rectangle, yshift=1.1cm, xshift=-2.3cm, anchor=east,font=\Large] {Category 1};
    \node (height) [rectangle, yshift=0.1cm, xshift=-2.3cm, anchor=east,font=\Large] {Category 2};
    \node (height) [rectangle, yshift=-0.9cm, xshift=-2.3cm, anchor=east,font=\Large] {Category 3};
    \node (height) [rectangle, yshift=-1.9cm, xshift=-2.3cm, anchor=east,font=\Large] {Category 4};
    \node (channels) [rectangle, yshift=-1.2cm, xshift=5.5cm,rotate=45,font=\Large] {Receiver};

\node[draw, single arrow,
              minimum height=25mm, minimum width=30mm,
              single arrow head extend=2mm,
              anchor=west] at (7.7, 0.2) {};

\node (layer2) [layer, xshift=12cm, yshift=0.2cm, yshift=0cm,font=\Large] {Layer $V$};
\draw [->] (layer2.east) -- (15.7, 0.2);

    \DrawCubes [step=10mm,thin]{17.99}{24}{2}{3}{0}{6}
    \node (width) [rectangle, yshift=-1cm, xshift=18.5cm,font=\Large] {Transmitter};
    \node (height) [rectangle, yshift=2.5cm, xshift=24cm, anchor=west,font=\Large] {Transmit power};
    \node (channels) [rectangle, xshift=14cm, yshift=0.8cm, xshift=9.5cm,rotate=45,font=\Large] {Receiver};
    
    \begin{scope}[canvas is xz plane at y=\YGridMax]
    \draw [fill=black!20!white]  (17.99,0) rectangle (18.99,1);
    \draw [fill=black!20!white]  (18.99,1) rectangle (19.99,2);
    \draw [fill=black!20!white]  (19.99,2) rectangle (20.99,3);
    \draw [fill=black!20!white]  (20.99,3) rectangle (21.99,4);
    \draw [fill=black!20!white]  (21.99,4) rectangle (22.99,5);
    \draw [fill=black!20!white]  (22.99,5) rectangle (23.99,6);
    \end{scope}
    
    \begin{scope}[canvas is xy plane at z=\ZGridMax]
    \draw [fill=black!20!white]  (22.99,2) rectangle (23.99,3);
    \end{scope}
    
    \begin{scope}[canvas is yz plane at x=\XGridMax]
    \draw [fill=black!20!white]  (2,5) rectangle (3,6);
    \end{scope}
\end{tikzpicture}}}
    \caption{Information processing in the \nn architecture.}
    \label{fig:info_processing}
\end{figure*}

The data flow along each layer of the network can be given by \eqref{eq:message-passing}. 
Here the node and edge features $(x_i,x_j,e_{j,i})$ are processed by edge processor $\phi$ and all the connected edges are aggregated to be passed on to the node processor by the aggregator $\bigoplus$.
Our architecture uses mean as the aggregator.
The node processor then updates the node feature based on the processed data. 
The processing in $\phi$ and $\gamma$ is illustrated in \autoref{fig:info_processing}.

Since each node $l$ knows the channel matrix $\mathbf{h}_{l,k}$, we denote it as $\mathbf{H} \in \mathbb{R}^{K \times K}$, where row $k$ and column $j$ is the channel gain $h_{kj}$.
Instead of optimizing $\mathbf{a}$ and $\mathbf{b}$ for a given $\mathbf{H}$, we define $\mathbf{a}=\alpha_\delta(\mathbf{H})$ where $\alpha_\delta$ is a \gls{gnn} as defined in \eqref{eq:message-passing} with parameters $\delta$ that maps $\mathbf{H}$ to $\mathbf{a}$.
Similarly, we also define $\mathbf{\ell} = \mathbf{b} - \mathbf{a} = \beta_\theta(\mathbf{H})$ where $\beta_\theta$ is a \gls{gnn} with parameter $\theta$ that maps $\mathbf{H}$ to the interval of the distribution that defines the transmit power.
Since the objective defined in \eqref{eq:objective2} is a function of $(\mathbf{H}, \mathbf{a}, \mathbf{b})$, we can write the objective $L$ as a function of $\delta$ and $\theta$.
\begin{equation}
    \label{eq:ml_objective}
    \max_{\delta, \theta} \sum_{\mathbf{H} \in \mathcal{D}} L(\mathbf{H}, \alpha_\delta(\mathbf{H}), \alpha_\delta(\mathbf{H}) + \beta_\theta(\mathbf{H}); \delta, \theta).
\end{equation}

We re-parameterize our objective according to \eqref{eq:ml} to optimize $\delta$ and $\theta$ in \eqref{eq:ml_objective} instead of optimizing $\mathbf{a}$ and $\mathbf{b}$ directly.
The training set $\mathcal{D}$ contains a large number of data samples and if the training data-set is general enough, it is expected that an optimized output can also be obtained for data sample $\mathbf{H}' \notin \mathcal{D}$.
This is expected because \gls{nn} follows the universal approximation theorem and with sufficient data, it can approximate the mapping very well.
With a trained $\alpha_\delta$ and a new channel realization $\mathbf{H}'$, we can compute $\alpha_\delta(\mathbf{H}')$ with low complexity and hence avoid the complicated iterative method described in Section~\ref{sec:rep}.
\section{Training and testing}
In this section we evaluate the proposed model on different scenarios of a \gls{cf-mmimo} network.
In every scenario, each \gls{ap} has $N=5$ antennas.
The number of \glspl{ap} considered in our simulations ranges from $5$ to $15$.
The number of \glspl{ue} ranges from $5$ to $15$ as well.
The \glspl{ap} are deployed $10~m$ above the \glspl{ue} in a grid. 
The \glspl{ue} are dropped according to uniform distribution in a $100 \times 100 ~m^2$ area.
The channels are modelled as $-30.5- 10 \times \alpha \times \log_{10}(d) $ using Rayleigh fading pathloss model~\cite{Bjornson2020} with carrier frequency of $2~\text{GHz}$, path-loss exponent is $3.67$, standard deviation of shadow fading $4~\text{dB}$ and noise power $-86\text{dBm}$.
The static power consumption is assumed to be $4\text{W}$ and the power amplifier inefficiency $\mu_i = 1$ for all $i$.
To evaluate the performance of the proposed model, we compare it with the state of the art \gls{sca} with Dinkelbach's transform in terms of performance metric and time.

A total of \num{10240} samples have been considered for training and, we divide the data into several smaller chunks of $1024$ samples each. 
We use ADAM optimizer with a \gls{lr} scheduler to update the parameters of \gls{gnn}. 
The configuration of \nn ~and the source code to reproduce the results can be found in the open source repository \url{https://gitlab.com/ichbinram/ee_cell_free}. 
All the hyper-parameters used for training are summarized in Table~\ref{tab:hyperparams}.

\begin{table}[ht]
\caption{Hyper-parameters used for $15$ \glspl{ap} and $15$ \glspl{ue} scenario}
\centering
\renewcommand{\arraystretch}{1.15}
\begin{tabular}{ll}
\toprule
\textbf{Parameters} & \textbf{\gls{gnn}} \\ 
\midrule
Initial learning rate & \num{0.001} \\ 
Final learning rate & \num{1e-7} \\
Framework & Unsupervised \\ 
Total data samples & \num{10240} \\
Batch size & \num{64} \\ 
Learning rate for $\kappa$ & \num{0.001}  \\ 
Iterations & \num{120}K \\
\bottomrule
\end{tabular}
\label{tab:hyperparams}
\end{table}

It is a common practice to normalize/standardize the data-set before being processed by a \gls{nn}.
In this case we noticed that transforming the data to have higher resolution yielded a better performance than standardization process.
This means that instead of using the well known standardization given as
\begin{equation}
    \label{eq:standardization}
    \hat{x}_i = (x_i-\mu)/\sigma,
\end{equation}
where $\hat{x}_i$ is the normalized data of sample $i$, $\mu$ is the mean and $\sigma$ is the standard deviation of the distribution that $x_i$ is sampled from,
we substitute $\mu$ with $\mu' = \mu - \epsilon$ and $\sigma$ with $\sigma' = \sigma -\epsilon'$ where $\epsilon ~\text{and} ~\epsilon'$ are small decrements.
This maps the dataset to a wide distribution with large values, meaning they are quite spread apart from each other.

\begin{figure}
    \centering 
    \begin{tikzpicture}

\definecolor{color0}{HTML}{004488}
\definecolor{color1}{HTML}{DDAA33}
\definecolor{color2}{HTML}{BB5566}

\begin{axis}[
width=.95\linewidth,
height=.22\textheight,
legend cell align={left},
legend style={
  fill opacity=0.8,
  draw opacity=1,
  text opacity=1,
  at={(0.97,0.03)},
  anchor=south east,
  draw=white!80!black
},
tick align=outside,
tick pos=left,
x grid style={white!69.0196078431373!black},
xlabel={Epoch},
xmajorgrids,
xmin=0, xmax=1.45e5,
xtick style={color=black},
y grid style={white!69.0196078431373!black},
ylabel={EE [Mbit/Joule]},
ymajorgrids,
ymin=0, ymax=9.93833293914795,
ytick style={color=black},
scaled x ticks=true,
xtick distance=2.5e4,
]
\addplot[] table[col sep=comma,header=true,x index=1,y index=2] {data/run-02-08-2023_13-20-00-tag-Training_ee.csv};
\addlegendentry{$\mu=2.5e-9, \sigma=6.9e-9$}
\addplot[color=color0] table[col sep=comma,header=true,x index=1,y index=2] {data/01-08-2023_16-19-47.csv};
\addlegendentry{$\mu'=1^{-11}, \sigma'=1^{-10}$}
\addplot[color=color1] table[col sep=comma,header=true,x index=1,y index=2] {data/run-03-08-2023_09-22-14-tag-Training_ee.csv};
\addlegendentry{$\mu'=1^{-12}, \sigma'=1^{-11}$}
\addplot[color=color2] table[col sep=comma,header=true,x index=1,y index=2] {data/run-05-08-2023_13-54-52-tag-Training_ee.csv};
\addlegendentry{$\mu'=1^{-13}, \sigma'=1^{-12}$}
\end{axis}
\end{tikzpicture}
    \caption{Improvement of objective with training for different cases of mean $\mu$ and standard deviation $\sigma$ for 15 \glspl{ap} and 15 \glspl{ue} scenario.}
    \label{fig:training}
\end{figure}
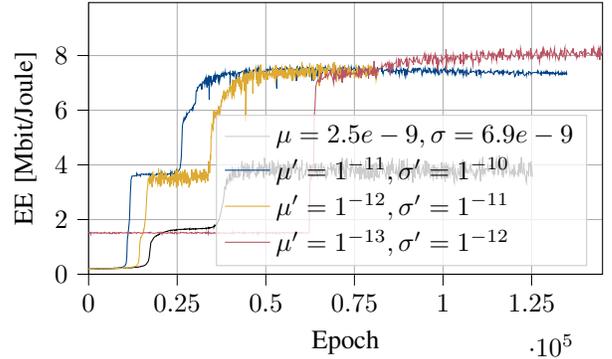

We believe that \gls{gnn} is able to better distinguish the data samples when they have higher resolution and map it to a better solution.
This can be seen as the increase in the performance of \gls{ee}.
This can be seen in Figure~\ref{fig:training}, 
which shows a comparison of the objective over $100$k epochs with different normalization values for $15$ \glspl{ap} and $15$ \glspl{ue}.
It can be observed that the model performs the worst for the standardization technique which is represented by the black curve, but the performance is increased when we convert the distribution into a non standard distribution.

It is also clear that decreasing the mean and standard deviation results in increased number of iterations for the model to converge.
This can be seen with the red curve which depicts the least mean and standard deviation, it can be seen being stuck in a local optima until \num{60}K iterations whereas the blue curve that is slightly lower than the actual mean and standard deviation of the training data-set already has achieved its best performance around \num{30}K iterations.

Throughout the process of training, \gls{csi} is provided to the \gls{gnn} in batches and we get the corresponding power allocation for this batch of data along with support of the distribution from which power is sampled. 
Using these input and output of the \gls{gnn}, we compute the loss function as explained in \autoref{subsec:non-convex}.
Finally, we compute the gradients of the loss function \wrt ~ parameters of \gls{gnn} and then update the parameters based on these gradients. 
We continue this process for several batches and this way, the \gls{gnn} learns to map power allocation to provided \gls{csi}.

For testing, we consider $1024$ samples for both \gls{sca} and \gls{gnn} and compare their average performance.
Each sample contains channel information of the \glspl{ue} and \glspl{ap} at different locations.
The channel data in the testing data set is mutually exclusive with the training set which means that \gls{gnn} has not been introduced to this data before.
It is observed that \gls{sca} takes considerably longer than the proposed method since it has to optimize for each of the data sample, whereas the parameters of \gls{gnn} are already optimized on a generalized dataset during training.
Therefore in the application phase, there is no need of complex optimization steps and  
All the tests have been carried out on AMD Ryzen \num{5} \num{5600}X \num{6}-Core Processor with \num{16} GB RAM. Table~\ref{tab:results} shows the performance comparison between the proposed\gls{gnn} and \gls{sca}.
\begin{table}[ht]
    \renewcommand{\arraystretch}{1.15}
    \centering
    \caption{Comparison of \gls{ee} [Mbit/Joule] performance and time [s] between \gls{sca} and \gls{gnn} on \num{1024} samples }
    \label{tab:results}
    \begin{tabularx}{\linewidth}{X X X}
    \toprule
    \textbf{Configuration} & \textbf{\gls{sca}} & \textbf{{\gls{gnn}}}\\
    \midrule
    $5$ \glspl{ap}, $10$ \glspl{ue} & \num{4.79} (\num{155.77}s) & \num{5.67} (\num{2.14}s)\\
    $10$ \glspl{ap}, $10$ \glspl{ue} & \num{5.33} (\num{198.26}s) & \num{5.93} (\num{1.94}s)\\
    $15$ \glspl{ap}, $15$ \glspl{ue} & \num{6.29} (\num{330.68}s) & \num{7.89} (\num{2.48}s)\\
    \bottomrule
    \end{tabularx}
\end{table}
It can be seen that the proposed approach can outperform the baseline in a fraction of the time required by \gls{sca}, resulting in better global \gls{ee} by reducing the resources required for computing the power allocation.
\begin{figure}
    \centering
    \includegraphics[width=\linewidth]{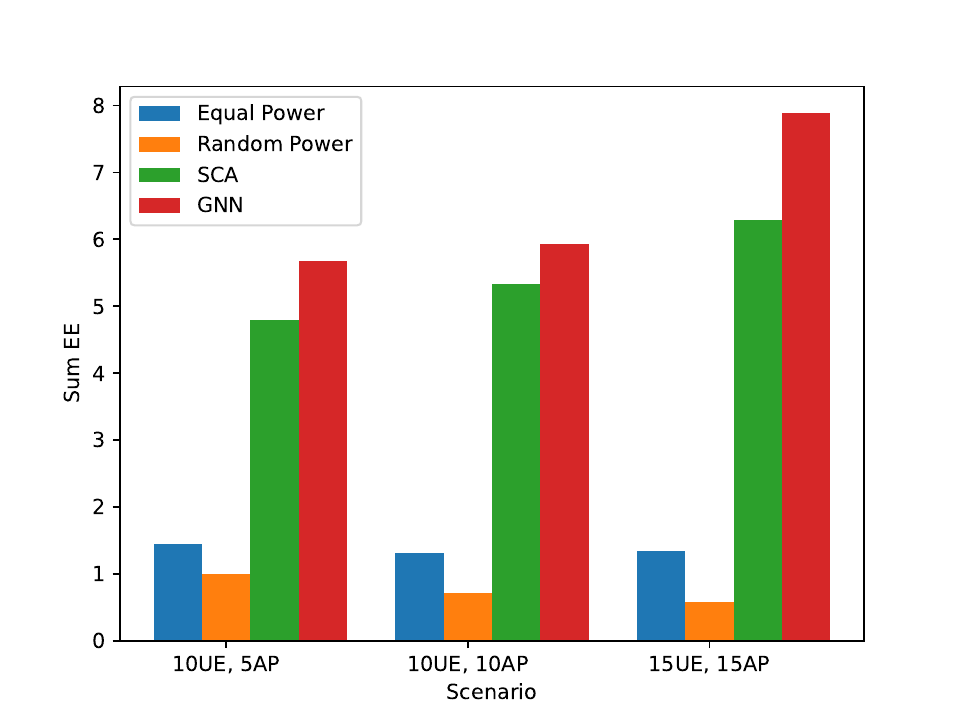}
    \caption{Comparison of sum EE [Mbits/Joule] performance for different schemes on 1024 samples}
    \label{fig:comparison}
\end{figure}

In \autoref{fig:comparison} we plot the \gls{ee} performance of the proposed approach with several benchmarks like random power allocation, equal power allocation and \gls{sca}.

\section{Conclusion}
\label{sec:conclusion}
\Gls{cf-mmimo} is a promising solution to future wireless communications, 
and \gls{ee} is a key objective of such communication systems.
Due to the non-convexity of the objective,
this problem has multiple local optima and cannot be solved with conventional convex optimization tools.
Although analytical methods perform considerably well it can not be applied to real-time application due to the high complexity.
Supervised learning framework have low complexity in the application phase but have an expensive data preparation routine during the training due to labeled data.
This paper presents an unsupervised machine learning solution to the problem.
We first consider a non-convex formulation of the loss function so that we can converge to the global optimum.
The support regularization guarantees the convergence.
Our second contribution is  to model the problem as a node level prediction task and design a dedicated \gls{gnn} architecture with nested SINRnet,
which encodes the domain knowledge of channels and realizes permutation-equivariance,
which is an inherent property of symmetry of multi-user power control problems.
The proposed architecture best fits the \gls{cf-mmimo} due to its distributed nature.
The proposed method achieves an \gls{ee} close to the solution obtained by the \gls{sca} algorithm in some cases outperforming it, approaching the global optimum.
In the future works, the approach to increase the scalability of training can be further investigated. 

\printbibliography

\end{document}